%% file: main.tex
\documentclass[10pt, a4paper]{article}

\usepackage[final]{lrec2026} 

\usepackage{times}
\usepackage[T1]{fontenc}
\usepackage[utf8]{inputenc}

\usepackage{microtype}

\usepackage{inconsolata}

\input{preamble}

\name{Yanis Perrin, Gilles Boulianne}
  
\address{
  Computer Research Institute of Montréal, Québec, Canada \\
  yanis.perrin@usherbrooke.ca, gilles.boulianne@crim.ca\\
}

\abstract{
Supervised training of speech recognition models requires access to transcribed audio data, which often is not possible due to confidentiality issues. Our approach to this problem is to generate synthetic audio from a text-only corpus using a state-of-the-art text-to-speech model with voice cloning capabilities. Our goal is to achieve automatic speech recognition (ASR) performance comparable to models trained on real data. We explore ways to optimize synthetic data generation through finetuning, filtering and evaluation, and its use for training an end-to-end encoder-decoder ASR model. Experiments were conducted using two datasets of spontaneous, conversational speech in Québec French. We show that improving data generation leads to large improvements in the final ASR system trained on synthetic data.
\\ \newline \Keywords{speech recognition, synthetic data, Québec French, Canadian French}}

\begin{document}
\maketitleabstract

\input{body_v2}

\nocite{*}
\section{Bibliographical References}

\bibliographystyle{lrec2026-natbib}
\bibliography{ref}

\appendix
\input{appendix}

\end{document}

%% file: preamble.tex
\usepackage[utf8]{inputenc} 
\usepackage[T1]{fontenc}    
\usepackage{hyperref}       
\usepackage{url}            
\usepackage{booktabs}       
\usepackage{amsfonts}       
\usepackage{nicefrac}       
\usepackage{microtype}      
\usepackage{lipsum}
\usepackage{fancyhdr}       
\usepackage{graphicx}       
\graphicspath{{media/}}     
\usepackage{multirow}
\usepackage{tabularx}
\usepackage{amsmath}        
\usepackage{comment}
\usepackage{tikz}
\usepackage{graphicx}
\usetikzlibrary{shapes.geometric, arrows.meta, positioning}
\usepackage{fixme}
\fxsetup{
    status=draft,
    author=,
    layout=inline,
    theme=color
}

\title{Towards Improved Speech Recognition through Optimized Synthetic Data Generation}

\author{
  Yanis Perrin \and Gilles Boulianne \\
  Computer Research Institute of Montréal, Québec, Canada \\
  \texttt{yanis.perrin@usherbrooke.ca \hfill gilles.boulianne@crim.ca} \\
}

\newcommand{\up}{$\uparrow$}
\newcommand{\devf}{$\text{dev}_\text{F}$}
\newcommand{\devm}{$\text{dev}_\text{M}$}
\newcommand{\testf}{$\text{test}_\text{F}$}
\newcommand{\testm}{$\text{test}_\text{M}$}
\newcommand{\dn}{$\downarrow$}
\newcommand{\WERRD}{\%WER$_\text{R}$\dn}

\newcommand{\WERR}{\%WER$_\text{R}$}
\newcommand{\WER}{\%WER }

\newcommand{\MOSU}{utMOS\up}
\newcommand{\MOS}{utMOS}
\newcommand{\bastr}{$\text{Bast}_\text{R}$}

\newcommand{\charbr}{$\text{Charb}_\text{R}$}

%% file: body_v2.tex
\section{Introduction}

Automatic Speech Recognition (ASR) systems have achieved remarkable performance in recent years, but this success has largely been contingent on access to vast amounts of transcribed audio data for training \cite{radford_robust_2023, zhang2023b, pratap2023a}. In specialized domains such as banking, healthcare and call centers, there exists a substantial deficit in such annotated data because of the highly confidential nature of the information in these recordings and the prohibitive costs associated with manual transcription. 

Concurrently, speech synthesis technology has witnessed significant advancements.  
Text-to-speech (TTS) systems can now generate increasingly natural and diverse speech from textual input, with state-of-the-art models capable of cloning voices with minimal reference audio \cite{casanova_xtts_2024, le_voicebox_2023}. 
Recent research has begun exploring the potential of using synthetically generated speech to supplement or replace real speech data in ASR training pipelines. However, the effectiveness of such approaches depends on multiple factors, including the quality and diversity of the synthetic data. Furthermore, the specific benefits of voice cloning capabilities in such systems for ASR training remain underexplored.
In this work, we investigate the feasibility and optimization of using synthetic speech data generated from voice-cloning TTS models to train robust speech recognition systems.

\subsection{Related Work}

Several studies have been conducted on the impact of synthetic speech to train ASR system over the past few years. Many of them use synthetic speech as part of data augmentation. \citet{laptev_you_2020} compared the performance of ASR models trained with semi-supervised learning or synthetic speech, and showed that results with synthetic speech were as good or better. 
\citet{fazel_synthasr_2021} demonstrated that even a small percentage of synthetic speech from the target dataset could greatly improve performance without any real target data. More recently, \citet{hilmes_effect_2024} investigated the effect of increasing TTS model complexity and varying speaker embeddings. They showed that the performance gap introduced by synthetic data depends on ASR architecture, with traditional models being more robust than modern end-to-end systems.
\citet{le_voicebox_2023} reports comparable performance of ASR models trained on either real or entirely synthetic speech.

Despite extensive work in the literature, most studies focus on clean, read speech, limiting their applicability to real-world scenarios with spontaneous speech and background noise. Furthermore, existing TTS models are primarily trained and evaluated on standard language variants, with limited attention given to underrepresented varieties such as Québec French (QF). 

This study addresses these gaps by investigating and optimizing synthetic speech generation in the context of spontaneous, domain-specific speech. Our key contributions are: (1) we adapt a general-purpose TTS model to the variety of French spoken in Québec, (2) we introduce methods to improve the effectiveness of synthetic data for ASR training through quality filtering mechanisms, (3) we simulate scenarios with varying levels of access to real and synthetic data, confirming the role of small amounts of real data, (4) we report results on two spontaneous QF datasets across different model architectures, and (5) we propose a method for generating textual content in scenarios where original textual data is unavailable.

\section{Methodology}

\begin{figure}[htb]
  \centering
  \includegraphics[width=1.0\linewidth]{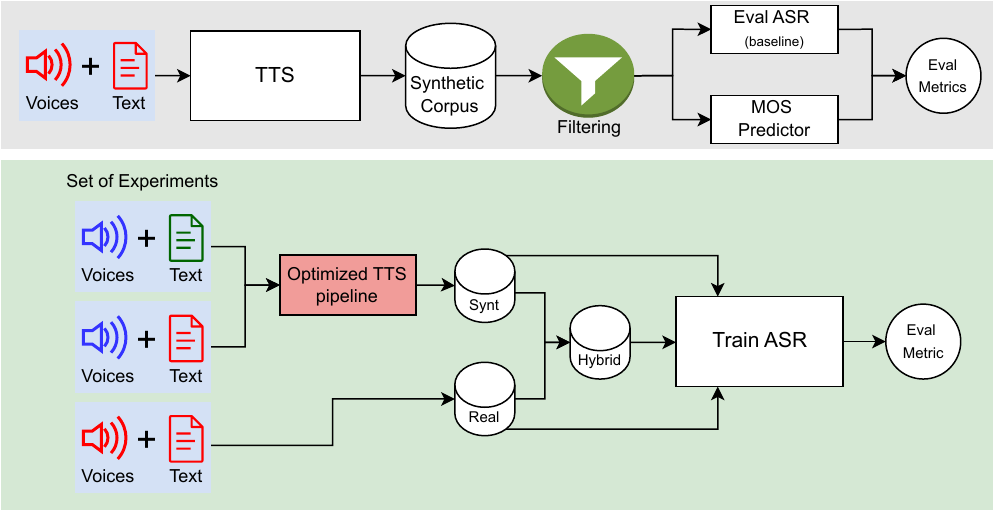}
  \caption{Top: Phase 1. Optimization of synthetic quality. Bottom: Phase 2. Speech recognition evaluation. Red indicates target domain data, blue out-of-domain data, and green, LLM-generated data. }
  \label{fig:methodology}
\end{figure}

To address this challenge, we use two real-world datasets containing spontaneous speech in QF, each providing both manual transcriptions and reference audio recordings. The smaller one is considered as the target domain and the largest as out of domain. Figure \ref{fig:methodology} illustrates the main phases of our methodology. In phase 1, the synthetic speech generation pipeline is optimized for quality, using
the target development set of transcriptions and voice references as input to the TTS. Each experiment  is a different configuration of finetuning and filtering method parameters. To measure the similarity between synthetic and real speech, we use as a proxy the word error rate (WER) of a speech recognizer trained exclusively on real speech (Eval ASR). We also evaluate the naturalness of the synthetic speech with utMOS \cite{saeki_utmos_2022} which approximates a subjective Mean Opinion Score of range 1 to 5. This work is the object of section \ref{sec:optimization_synth}.

In phase 2, we use the optimized TTS pipeline to generate a synthetic speech dataset and evaluate its effectiveness for training ASR models, alone or in combination with real data. The evaluation metric is word error rate, measured on real data, of the ASR trained on synthetic data. As described in section \ref{sec:asr_evaluation}, we explore gradually more complex inputs to the TTS pipeline, from target domain text transcriptions and voice samples, to out-of-domain voice samples, and finally with LLM-generated text and out-of-domain voice samples.

\subsection{Datasets}

While most ASR research on synthetic speech uses widely-available English-language benchmarks such as LibriSpeech \cite{panayotov_librispeech_2015}, our work focuses on QF speech. We conduct our experiments on CommissionsQC \cite{serrand2025a}, which includes Bast and Charb corpora, derived from Québec governmental inquiry commissions. 
These datasets present real-world acoustic conditions including background noise and microphone variations, spontaneous speech with hesitations and disfluencies, and distinct QF dialectal features that differ significantly from standard benchmarks. Bast and Charb contain 85.7 and 724 hours of speech, respectively. More details are provided in Table~\ref{tab:corpus_stat} of Appendix \ref{sec:app_tables}.

\subsection{Baseline results}

The performance of the evaluation ASR model when trained and tested on real data is listed in Table~\ref{tab:baseline_asr}. Additional details regarding the model architecture and training configuration are provided in Section \ref{sec:asr_evaluation}.

\begin{table}[htb]
    \centering
    \begin{tabular}{llrrrr}
    \toprule
    & & \multicolumn{4}{c}{\WERR} \\\cmidrule{3-6}
    Train & Test & $\text{dev}_\text{F}$ & $\text{dev}_\text{M}$ & $\text{test}_\text{F}$ & $\text{test}_\text{M}$ \\
    \midrule
    \bastr & \bastr & 13.8 & 13.5 & 12.0 & 16.3  \\     \bastr & \charbr & 22.1 & 24.4 & 26.8 & 24.3  \\
    \charbr & \charbr &  7.4 & 8.3 & 6.1 & 9.4  \\
    \charbr & \bastr  &  7.6 & 8.3 & 8.0 & 9.2   \\
    \bottomrule
    \end{tabular}
    \caption{Baseline results on real male and female development and test sets: {\%WER$_\text{R}$} refers to WER\ evaluated with a model trained on real data. $\text{Bast}_{\text{R}}$ and $\text{Charb}_{\text{R}}$ stand for the original corpora of Bast and Charb.}
    \label{tab:baseline_asr}
\end{table}

\subsection{Text-to-Speech model}

For speech synthesis, we selected XTTS-V2\footnote{\url{https://github.com/coqui-ai/TTS}.} \cite{casanova_xtts_2024}, a publicly available TTS that meets 3 primary criteria: (1) the capacity to synthesize speech from a given text while preserving the linguistic content and reproducing the acoustic characteristics of a short voice sample; (2) support for high-quality French speech synthesis; and (3) the capacity to be finetuned for domain-specific adaptation. 

\section{Optimization of synthetic quality}
\label{sec:optimization_synth}

Our objective is to create synthetic utterances that closely approximate the acoustic and linguistic characteristics of our authentic QF speech corpora. 
In our experiments, the Bast corpus is used as the target domain for synthetic replication due to its relatively limited size, which enables more efficient generation of synthetic datasets and facilitates a broader range of experimental conditions. In this setup, the Charb corpus serves as an out-of-domain source to provide reference voices for synthesizing speech with Bast text.

\subsection{Finetuning for Québec French}
\label{finetuning}

For French, most of XTTS training data is from Common Voice \cite{casanova_xtts_2024}, which we estimate contains less than 5\% QF. Preliminary experiments using the original model to synthesize QF yielded suboptimal results. To address this dialectal mismatch, we finetuned the XTTS-V2 model using a 12 hours subset of the Charb corpus. We first restricted utterance duration to a range of 5-15 seconds to avoid untypical lengths.
Then, we eliminated unusually fast or slow utterances, more specifically those in the shortest or longest 10\% of the text length distribution for a given duration.
We cyclically iterated through all the speakers and picked one utterance per speaker in each pass, to maintain the same speaker distribution. 

Finetuning followed the Coqui github recipe\footnote{\url{https://github.com/coqui-ai/TTS}}, with the default hyperparameters. We further tuned the learning rate and number of training steps hyperparameters. A learning rate of 5e-6 and 60k training steps yielded the lowest WER and highest \MOS\  on both development and test. See Table~\ref{tab:hyperparameters_TTS} in Appendix \ref{sec:app_tables} for more details.

\subsection{Filtering}
After hyperparameter optimization, we observed that our synthetically generated utterances still presented various artifacts including hallucinations, disfluencies, and inappropriate pauses, that could potentially degrade ASR training effectiveness. To reduce these artifacts, we developed filtering methods to identify and regenerate problematic samples.

Our first intuition was that significant differences in duration between synthetic and original speech may indicate a problem. 
For each utterance, we computed the ratio of synthetic duration to reference duration.
When this ratio falls outside a predetermined acceptable range, the utterance is flagged for regeneration. 
Through empirical analysis of previous generation attempts, we identified thresholds that effectively separated clean utterances from those with artifacts.

Such duration-based filtering yielded only marginal improvements in WER. Our analysis indicates that it is primarily effective in identifying highly degraded utterances with large alterations of the temporal structure of speech, which are relatively infrequent in our finetuned TTS system. More detailed results are available in Table ~\ref{tab:duration_TTS} of Appendix \ref{sec:app_tables}.

\paragraph{Generator-Verifier}

Duration filtering approach does not explicitly evaluate linguistic content fidelity. To address this limitation, we propose leveraging a robust ASR model for content validation. The assumption is that a strong ASR system failing to correctly transcribe a given audio sample means that it likely contains noise, distortions, or linguistic content that does not match the TTS input text.

We designed a generation pipeline composed of a \textbf{Generator} and a \textbf{Verifier}. The generator, which may be any TTS model, is responsible for producing synthetic utterances. The verifier, implemented as a pretrained ASR model, transcribes each synthetic utterance, and its transcription is compared to the original transcription using the WER metric. If the computed WER exceeds a predefined threshold, the sample is rejected and regenerated. For computational efficiency, we allow a maximum of ten generation attempts per utterance. 

\begin{figure}[htb]
  \centering
  \includegraphics[width=1.0\linewidth]{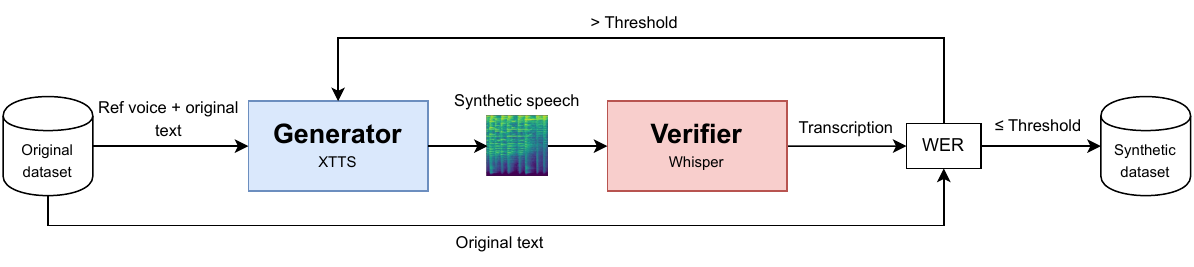}
  \caption{Generator-Verifier based filtering }
  \label{fig:pipeline-drawio}
\end{figure}

As the verifier component, we use Whisper \cite{radford_robust_2023} for easy reproducibility. 
We use the large-v3 model in its optimized version faster-whisper\footnote{\url{https://github.com/SYSTRAN/faster-whisper}} as it offers correct transcription performance on QF and less hallucinations than the medium size.  The generator component is based on our finetuned version of the XTTS model. The overall architecture of the proposed filtering pipeline is illustrated in Figure~\ref{fig:pipeline-drawio}.

Table~\ref{tab:wer_TTS} illustrates the impact of this filtering. While duration-based filtering achieved a maximum WER reduction of 0.2\% absolute on the development set, the proposed verifier-based filtering improves this reduction to 2.4\%, and further achieves a 3.0\% reduction on the test set. 
By directly evaluating the linguistic content rather than relying on indirect measures, G-V filtering appears more effective at detecting hallucinations and selecting higher-quality synthetic data.
\begin{table}[htb]
    \centering
    \begin{tabular}{lrrrrr}
    \toprule
     & \multicolumn{2}{c}{\WERRD} & \multicolumn{2}{c}{\MOSU}\\
    Threshold & $\text{dev}$ & $\text{test}$ & $\text{dev}$ & $\text{test}$  \\
    \midrule
    None  &  21.9 & 22.0 & 2.34 & 2.25 \\
    0.30 &  20.2 & 20.0 & \textbf{2.36} & 2.25  \\ 
    0.25 & 20.1 & 19.3 & \textbf{2.36} & \textbf{2.26} \\
    0.20 &  \textbf{19.5} & \textbf{19.0} & 2.35 & \textbf{2.26}  \\ 
    
    \bottomrule
    \end{tabular}
    \caption{G-V filtering of synth: {\%WER$_\text{R}$} and utMOS evaluation for different values of rejection threshold. \textit{None} indicates that no filtering was applied.}
    \label{tab:wer_TTS}
\end{table}

\subsection{Generation temperature}

We generated synthetic datasets with temperature of 0.1, 0.65 and 1.3. Lower temperatures produced higher-quality speech, as evidenced by reduced \WERR\ and \MOS\ values. However, at the lowest temperature of 0.1, we observed a lack of prosodic and acoustic diversity. We generated training sets with a temperature of 0.65 and found that quality metrics decreased, but trained ASR model performed better. So we adopted a temperature of 0.65 for generating all synthetic training sets. More detailed results can be found in Table \ref{tab:temperature_01_asr} of Appendix \ref{sec:app_tables}. 

\subsection{Combining improvements}

Table~\ref{tab:ablation_tts} shows a successive refinements of the methods cumulate into a large improvement in our speech synthesis quality metrics.
Finetuning for QF and G-V filtering are the most important individual contributions to the overall improvement.

\begin{table}[htb]
    \centering
    \begin{tabular}{lrrrr}
    \toprule
    TTS & \multicolumn{2}{c}{\WERRD} & \multicolumn{2}{c}{\MOSU}  \\ \cmidrule{2-5}
    condition   & $\text{dev}$ & $\text{test}$ & $\text{dev}$ & $\text{test}$ \\
    \midrule
    Initial xtts                &  30.8  & 29.4     & 2.04     &   2.01 \\
    Finetuning for QF & 24.2 & 24.3 & 2.27 & 2.19  \\
    \quad Hyperparams  & 21.9 & 22.0 & 2.34 & 2.25 \\
    \quad G-V filtering             & 19.5 & 19.0 & 2.35 & 2.26  \\
    \quad Temperature    & 17.7 & 17.5 & 2.42 & 2.32 \\     Charb voice  & 17.2 &  17.0  & 2.50 & 2.55 \\
    \bottomrule
    \end{tabular}
    \caption{Progress in quality metrics of synthetic Bast development and test sets.}
    \label{tab:ablation_tts}
\end{table}

\section{Speech recognition evaluation}
\label{sec:asr_evaluation}

In order to evaluate the generated synthetic data usefulness to train a speech recognition model, we generate multiple training datasets varying in size and composition.

\subsection{Training data generation}
\label{sec:training_data_gen}

We create purely synthetic datasets of 90 and 360 hours to assess the effect of synthetic dataset size. For evaluating the impact of combining synthetic and real speech, we also construct hybrid datasets combining synthetic (S) and real (R) speech, each totalling 360 hours but with different real-to-synthetic ratios: 350h S / 10h R, 330h S / 30h R, and 300h S / 60h R. To explore larger-scale configurations, we also produce hybrid datasets of 720 and 770 hours, composed of 710h S / 10h R and 710h S / 60h R, respectively, aligning with the scale of the Charb corpus. 

\paragraph{Reference Voices} 

To simulate a scenario in which we don’t have access to original recordings, we selected reference voices from the out-of-domain Charb corpus.
For each Charb speaker, 
a list of utterances is drawn up, with duration of 8 to 12 seconds, and text length in the $10^{th}$ to $90^{th}$ percentile of the text length distribution, and one is picked at random. If for a given speaker, no utterance meets the conditions, a "best of the bad" utterances is chosen as the one closest to acceptable.
For Charb there were 342 speakers in training. One utterance matched for 248 speakers, and the remaining 94 speakers were assigned the "best of their bad" utterances. Thus we end up with one audio utterance to represent each Charb speaker.

\paragraph{Synthetic corpus} During generation, each original transcription from Bast is randomly selected (with replacement) and paired with a randomly selected speaker from the Charb corpus. Pairings are constrained to match the gender and dataset partition (train, development, or test) of the original Bast speaker to preserve consistency. Each text/audio pair appears only once. This strategy prevents speaker overlap across training, development, and test sets, when creating datasets that exceed the size of the original corpus.

\paragraph{Real and hybrid corpus} Hybrid corpora are constructed by randomly sampling real and synthetic data from their respective original and generated corpora. To maximize speaker diversity, utterances are selected by cyclically iterating over the speakers in each corpus without replacement. For consistency and comparability across experiments, a nested data-sharing approach is employed: each smaller corpus is a strict subset of the larger ones. The same nesting is applied to real datasets. 

\paragraph{Text generation}

To simulate a scenario in which we don’t have access to original recordings and transcriptions, we generate texts intended to closely resemble those from the Bast corpus, based on general information and written documents. The texts are then input for TTS generation, while reference voices are sampled from the Charb corpus. Our target is to generate 50k utterances, as in the preprocessed Bast corpus.

To achieve this, we used OpenAI's GPT-4o as our prompt engineer to generate instructions for our corpus generator GPT-4o-mini, a lightweight language model optimized for faster inference. To replicate the original text distribution, we used a structured prompting strategy. A base global prompt (created by GPT-4o) provided the model with contextual grounding. Then, five specialized sub-prompts were applied iteratively, with one sub-prompt introduced every four generations. These sub-prompts were based on a qualitative analysis of the model's initial generations, and designed to introduce themes and content underrepresented in early outputs. The prompts appear in Appendix~\ref{sec:app_prompts}.

\paragraph{ASR training setup}

The ASR model we use is an end-to-end transformer-based encoder-decoder architecture with 29.4M trainable parameters\footnote{\url{https://github.com/espnet/espnet/blob/master/egs2/librispeech_100/asr1/conf/tuning/train_asr_transformer_win400_hop160_ctc0.3_lr2e-3_warmup15k_timemask5_amp_no-deterministic.yaml}} implemented and deployed using the ESPnet \cite{watanabe_espnet_2018} toolkit. We choose this model because it is both fast to train and achieves a reasonable recognition rate on LibriSpeech\_100, which is comparable in size to Bast. 
Audio data were sampled at 16 kHz. For feature extraction, we applied a sliding window of 25 ms (400 samples) with 40\% overlap between consecutive frames. We excluded utterances shorter than 2 seconds and longer than 30 seconds. Tokenization uses Byte-Pair Encoding (BPE) with a vocabulary size of 5,000 subword units. 

The ASR model was trained from scratch for 70 epochs using the Adam optimizer \cite{kingma2017adammethodstochasticoptimization} with weight decay regularization (coefficient: 1e-06). We used a warm-up strategy with 9,000 steps to gradually increase the learning rate to 4e-03. The loss function used a weighted combination of CTC loss (0.3) and label smoothed cross-entropy loss (0.7) (\cite{boyer2022studytransducerbasedendtoend} to optimize both frame-level alignments and sequence-level predictions. To improve generalization, dropout is applied systematically throughout the network at a consistent rate of 0.1. All training was conducted on an NVIDIA A40 GPU.
We used hybrid decoding with beam search and a beam width of 20,  and combined CTC and attention scores with weights of 0.3 and 0.7 respectively.

\subsection{Results}
The WER of models trained on fully synthetic, fully real and hybrid datasets is evaluated on the real development and test sets. Results are presented in Table~\ref{tab:real_vs_synth_asr}. In this table, the number next to the dataset indicates the total duration in hours. For hybrid corpora, the first value corresponds to synthetic data and the second to real data. 

\begin{table}[htb]
    \centering
    \begin{tabular}{llcc}
    \toprule
     &  & \multicolumn{2}{c}{\%WER\dn} \\ \cmidrule{3-4}
    Train & Eval & $\text{dev}$ & $\text{test}$ \\
    \midrule
    $\text{Bast}_\text{R-10}$ & \bastr & 80.8 & 82.1 \\
    $\text{Bast}_\text{S-90}$ & \bastr & 26.9 & 28.2 \\     $\text{Bast}_\text{S-360}$ & \bastr & 26.2 & 27.1   \\
    $\text{Bast}_\text{Mix-350/10}$ & \bastr & 24.5 & 25.2 \\
    $\text{Bast}_\text{Mix-710/10}$ & \bastr & 22.4 & 22.8 \\
    $\text{Bast}_\text{Mix-330/30}$ & \bastr & 21.7 & 22.2   \\
    $\text{Bast}_\text{Mix-300/60}$ & \bastr & 20.8 & 21.3 \\
    $\text{Bast}_\text{Mix-710/60}$ & \bastr & 20.4 & 20.5 \\
    $\text{Bast}_\text{R}$ & \bastr & \textbf{13.6} & \textbf{14.2} \\
    \bottomrule
    \end{tabular}
    \caption{\%WER when training on real (R), synthetic (S), and hybrid (Mix) datasets. Evaluation on real data.}
    \label{tab:real_vs_synth_asr}
\end{table}

The model trained exclusively on 10 hours of real data, $\text{Bast}_\text{R-10}$, fails to converge. Both purely synthetic datasets outperformed this 10-hour real data baseline, but remain worse than the baseline model trained on the full real dataset $\text{Bast}_\text{R}$. Modest improvements of 0.7\%-1.1\% are observed on the development and test sets, respectively, between the two synthetic datasets as they increase from 90 to 360 hours, with diminishing returns. A performance gap of 13\%-14\% absolute remains between synthetic and real speech training.  

With hybrid datasets, we observe a positive correlation between the proportion of real data and ASR model performance. 
With 10 hours of real data, increasing synthetic data from 350 to 710 hours yielded reductions WER of 2.1\% and 2.4\% absolute on the development and test sets. However, this benefit diminishes as the proportion of real data increases: comparing $\text{Bast}_\text{Mix-300/60}$ and $\text{Bast}_\text{Mix-710/60}$, only marginal WER reductions of only 0.4\% and 0.8\% absolute are seen on the development and test sets, respectively. These findings indicate that synthetic data augmentation provides substantial benefits in data-scarce scenarios ($\leq$ 10 hours real data) but offers diminishing returns when sufficient real data is available. More detailed results are available in Table~\ref{tab:overview_asr} in Appendix \ref{sec:app_tables}.

Results on datasets created from the LLM-generated text are listed in Table~\ref{tab:gentext_asr}. They indicate that synthetic text distribution remains significantly different from that of the original corpus, as evidenced by the high WER observed in 1st row compared to the 2nd row. However, when 10 hours of real audio are added, 
WER is reduced by half, yielding performance comparable to and even better that of the 90-hour synthetic-only condition. 

\begin{table}[htb]
    \centering
    \begin{tabular}{lcc}
    \toprule
     & \%WER\dn & \%WER\dn  \\
    Train & $\text{dev}$ & $\text{test}$ \\
    \midrule
     $\text{Gen-text}_\text{S-360}$ & 50.9 & 54.2 \\
     $\text{Bast}_\text{S-90}$ & 26.9 & 28.2   \\
    $\text{Gen-text}_\text{Mix-350/10}$ & 26.2 & 27.7 \\ 
         $\text{Bast}_\text{Mix-350/10}$ & 24.5 & 25.2 \\
             \bottomrule
    \end{tabular}
    \caption{Comparison of \%WER when training with synthetic speech generated using real or generated transcriptions (\text{Gen-text}), and evaluating on real data.}
    \label{tab:gentext_asr}
\end{table}

\subsection{Pretrained model finetuning}

To assess the potential benefit of synthetic data for adapting pretrained models to task-specific domains, we explored finetuning the medium variant of the Whisper model with Low-Rank Adaptation (LoRA) \cite{hu2021loralowrankadaptationlarge}. We followed the configuration default values provided in the AISHELL-1 \cite{bu2017aishell1opensourcemandarinspeech} ESPnet recipe\footnote{\url{https://github.com/espnet/espnet/blob/master/egs2/aishell/asr1/conf/tuning/train_asr_whisper_medium_lora_finetune.yaml}}. 
A series of experiments were conducted under this setup, with the corresponding results summarized in Table~\ref{tab:whisper_finetuning_results}.

\begin{table}[htb]
    \centering
    \begin{tabular}{llrr}
    \toprule
     & &  \multicolumn{2}{c}{\%WER\dn} \\
    Status & Train & $\text{dev}$ & $\text{test}$ \\
    \midrule
    Pretrained & None & 11.4 & 10.2 \\  
     Finetuned &$\text{Bast}_\text{S-360}$ & 10.5 & 14.8   \\
     Finetuned &$\text{Bast}_\text{Mix-350/10}$ & 7.2 & 9.5 \\ 
    Finetuned & $\text{Bast}_\text{R-10}$ & 6.5 & 8.0 \\
     Finetuned &$\text{Bast}_\text{R}$ & 5.7 & 7.1 \\ 
    \bottomrule
    \end{tabular}
    \caption{\%WER after finetuning Whisper-medium with various combinations of real and synthetic speech. Results reported on real development and test sets.}
    \label{tab:whisper_finetuning_results}
\end{table}

Pretrained Whisper model demonstrates better performance compared to our baseline Transformer trained from scratch, achieving lower WER without domain-specific finetuning. 
 
Finetuning further improves the results.
 
These improvements continue as the amount of real data increases, reaching best performance with the complete original dataset.   
Hybrid finetuning using 350 hours of synthetic data and just 10 hours of real data yields stronger results than synthetic data alone,
the best results obtained with predominantly synthetic training data
but still not surpassing those achieved by finetuning with only 10 hours of real data. This suggests that even limited real data is more effective than a large volume of lower-quality synthetic data which may interfere with prelearned representations. 

\section{Conclusion} 
We explored the effectiveness of generating synthetic speech datasets for ASR training, using data from two Québec commissions, across multiple scenarios of varying data availability, including a simulation with no initial real data. We proposed methods to improve synthetic speech quality and demonstrated that optimizing the generation processes enhanced synthetic data quality. Our results confirm the importance of incorporating small amounts of real target data to improve recognition model performance. Despite our optimizations, a quality gap between real and synthetic data persists, underscoring the need for further refinement in speech data generation approaches. 

\section{Limitations}
In this study, we used XTTS as our only synthesis model, which may have limited the diversity of the resulting generated data. We considered conducting experiments using the model proposed in \cite{le_voicebox_2023} due to its promising performance on LibriSpeech, but its unavailability to the research community precluded its use in our study. Investigating the use of multiple Text-to-Speech models with varying architectures should be explored to enhance acoustic and linguistic variability and reduce reliance on the limitations of a single-synthesis system.
Additionally, alternative methods for assessing the similarity between synthetic and real data, such as embedding-based clustering techniques, could be explored. 

\section{Ethical Considerations}
The approach we propose involves cloning voices from audio samples, raising ethical considerations regarding data privacy and consent. We acknowledge the importance of ensuring informed consent from all individuals whose voice samples are used for model training, particularly given the potential applications in sensitive domains discussed in the Introduction section. 
In this work, the approach requires only a single voice sample per speaker to produce synthetic speech, which reduces the need to access additional customer recordings and ensures greater confidentiality. 

\section{Acknowledgements}

We acknowledge the support of the Natural Sciences and Engineering Research Council of Canada (NSERC) for this work, and would also like to thank Ministry of Economy and Innovation (MEI) of the Government of Québec for its continued support.

%% file: appendix.tex
\section{Prompts for Text Generation}
\label{sec:app_prompts}

\begin{figure}[ht]
  \centering
  \includegraphics[width=1.0\linewidth]{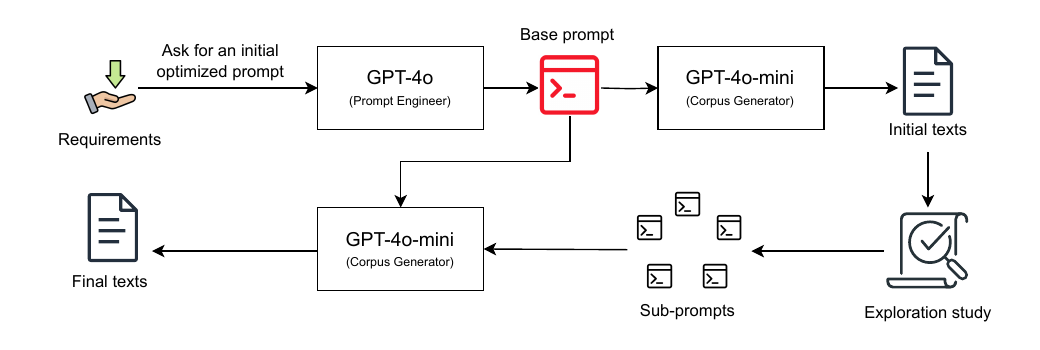}
  \caption{Synthetic text generation pipeline }
  \label{fig:generation-text-figure}
\end{figure}

We initially used GPT-4o to generate a base prompt, tailored to our predefined requirements. This prompt was then provided to GPT-4o-mini to produce the first batch of synthetic texts. Upon analyzing these initial outputs, we identified limitations in relevance, which led us to develop five targeted sub-prompts to guide the generation more effectively. The process was restarted using the base prompt followed by the sub-prompts in an iterative manner, every four generations, until the desired number of utterances was produced. The base prompt and the five sub-prompts are provided below.

\subsection{Base Prompt}
\label{subsubsec:initial_prompt}

\vspace{0.2em}
\paragraph{Objective:}
Orient the model and provide initial context for text generation.

\vspace{0.2em}
\paragraph{Prompt Content:}

\begin{quote}
\textbf{Contexte :} Vous devez créer des transcriptions de dialogues pour une commission d'enquête au Québec sur le processus de nomination judiciaire. Le sujet inclut l'influence potentielle de tiers et les critères de sélection des juges des cours municipales et des membres du Tribunal administratif du Québec. Les échanges se font en français québécois, reflétant le registre et les nuances culturelles de cette langue.

\vspace{0.2em}
\textbf{Instructions :}
\begin{enumerate}
    \item Créez des dialogues réalistes et centrés sur le sujet principal, permettant de couvrir différentes dimensions du processus de nomination judiciaire.
    
    \item Intégrez des détails concrets dans les échanges, tels que des dates précises, années, numéros d'articles de loi, références de documents déposés, noms de villes, lieux, partis politiques, montants financiers, et procédures spécifiques.
    
    \item Variez les intervenants, incluant :
    \begin{itemize}
        \item Le président de la commission
        \item Des experts juridiques
        \item Des témoins
        \item Des citoyens ou parties prenantes
    \end{itemize}
    
    \item Indiquez seulement le prénom et le nom de chaque intervenant, suivi de leur texte dans la transcription.
    
    \item Assurez-vous que les échanges incluent aussi des réponses très courtes et naturelles de quelques mots, pour refléter des dialogues réalistes.
    
    \item Gardez les conversations pertinentes au thème principal, tout en explorant des sous-thèmes comme :
    \begin{itemize}
        \item Critères de qualification et éthique
        \item Influence politique et externe
        \item Transparence et responsabilité
    \end{itemize}
    
    \item Structurez le texte pour qu'il reflète un échange continu sans introductions ni conclusions formelles.
    
    \item Adaptez le registre de langue selon les participants :
    \begin{itemize}
        \item Un ton formel pour le président et les experts
        \item Un langage plus spontané pour les témoins et citoyens
    \end{itemize}
    
    \item Intégrez des variations émotionnelles et subtilités culturelles pour rendre les échanges vivants, toujours sous forme de dialogue.
    
    \item Respectez les conventions d'écriture :
    \begin{itemize}
        \item Évitez les abréviations (ex. ``Monsieur'' au lieu de ``M.'')
        \item Écrivez les nombres en toutes lettres
        \item Maintenez une mise en page claire
    \end{itemize}
\end{enumerate}

\vspace{0.2em}
\textbf{Exemple de format à respecter :}
\begin{quote}
\textit{Président :} ``Quelle est votre opinion sur l'indépendance judiciaire dans le processus de nomination ?''

\textit{Madame Tremblay :} ``Je crois fermement que le processus doit être transparent et exempt de toute influence...''

\textit{Monsieur Giguère :} ``C'est exact.''
\end{quote}
\end{quote}

\subsection{Continuation of dialogues}
\label{subsubsec:continuation of dialogues}

\vspace{0.2em}
\paragraph{Objective:} 
Encourages the model to extend previously generated dialogues for greater coherence.

\vspace{0.2em}
\paragraph{Prompt Content:}

\begin{quote}

\vspace{0.2em}
\textbf{Instructions :}
\begin{enumerate}
    \item Continuez les dialogues en gardant bien à l'esprit que le sujet principal est le processus de nomination des juges au Québec.
    \item Maintenez le flux de la conversation, en développant progressivement les idées et les arguments de manière approfondie avec un vocabulaire approprié à une commission d'enquête.
    \item Continuez de structurer le texte pour qu’il reflète un échange continu sans introductions ni conclusions, typique d'une commission d'enquête.
    \item Produisez une variété de dialogues, alternant entre des échanges courts et d'autres plus approfondis.
    \item Explorez de nouveaux aspects du sujet si le précédent est épuisé, et introduisez de nouveaux participants au besoin, sans forcer des conclusions précipitées.
\end{enumerate}
\end{quote}

\subsection{Development of Facts}
\label{subsubsec:development of facts}

\vspace{0.2em}
\paragraph{Objective:} 
Promotes the inclusion of specific factual elements such as dates, places, political parties, and legal references.

\vspace{0.2em}
\paragraph{Prompt Content:}

\begin{quote}
\vspace{0.2em}
\textbf{Instructions :}
\begin{enumerate}
    \item Continuez le dialogue en vous concentrant sur le développement de faits passés, en questionnant les témoins et les experts à ce sujet.
    \item Concentrez-vous sur la génération de témoignages détaillés et de dialogues investigatifs, en explorant des sujets spécifiques avec des faits concrets.
    \item Formulez des questions précises sur des événements antérieurs liés au processus de nomination des juges, sollicitant des détails et clarifications.
    \item Maintenez un ton investigatif, s'assurant que la conversation demeure engagée et directement liée aux faits examinés.
\end{enumerate}
\end{quote}

\subsection{Familiar Language}
\label{subsubsec:familiar language}

\vspace{0.2em}
\paragraph{Objective:} 
Guides the model to produce utterances in informal Quebec French, capturing regional linguistic nuances.

\vspace{0.2em}
\paragraph{Prompt Content:}

\begin{quote}
\vspace{0.2em}
\textbf{Instructions :}
\begin{enumerate}
    \item Continuez le dialogue en intégrant des échanges plus spontanés et familiers, surtout pour les témoins et les citoyens.
    \item  Utilisez un langage qui reflète les nuances et le ton informel du français québécois, en vous éloignant d’un registre trop formel.
    \item Encouragez les participants à répondre avec des expressions courantes et des réactions naturelles pour rendre les interactions plus authentiques.
    \item Maintenez le focus général sur le processus de nomination des juges, mais avec une approche plus décontractée de la conversation.
\end{enumerate}
\end{quote}

\subsection{Short Answers}
\label{sec:short_answers}

\vspace{0.2em}
\paragraph{Objective:} 
Constrains the model to generate concise responses, preventing overly long or verbose texts.

\vspace{0.2em}
\paragraph{Prompt Content:}

\begin{quote}
\vspace{0.2em}
\textbf{Instructions :}
\begin{enumerate}
    \item Continuez le dialogue en encourageant des réponses très courtes et des échanges rapides entre les participants incluant des réponses simples de quelques mots.
    \item Introduisez de nouvelles voix ou perspectives si nécessaire, mais maintenez un échange très rapide et concis.
    \item Assurez-vous que le ton et le style restent conformes au français québécois.
    \item Assurez-vous que les interactions restent centrées sur le processus de nomination des juges au Québec, tout en intégrant divers points de vue.
\end{enumerate}
\end{quote}

\subsection{Exploration of Unrelated Topics}
\label{subsubsec:short answers}

\vspace{0.2em}
\paragraph{Objective:} 
Orient the model to explore other more distant subjects but which remain linked to the process of appointing judges.

\vspace{0.2em}
\paragraph{Prompt Content:}

\begin{quote}
\vspace{0.2em}
\textbf{Instructions :}
\begin{enumerate}
    \item Poursuivez le dialogue en explorant des sujets plus éloignés mais toujours liés au processus de nomination des juges.
    \item  Continuez de structurer le texte pour qu’il reflète un échange continu sans introductions ni conclusions.
    \item Produisez cette fois des échanges plus long tout en alternant avec quelques échanges courts.
    \item Maintenez un style engageant et informatif tout en explorant ces nouvelles dimensions.
\end{enumerate}
\end{quote}

\section{Tables}
\label{sec:app_tables}

\begin{table}[htb]
    \centering
\begin{tabular}{llrrrr}
    \toprule
     &  & \multicolumn{2}{c}{\WERR} & \multicolumn{2}{c}{\MOS} \\
    LR & Steps & $\text{dev}$ & $\text{test}$ & $\text{dev}$ & $\text{test}$ \\
    \midrule
    None & None  &  30.8 & 29.4 & 2.04 & 2.01   \\
    1e-06 & 15k  & 29.7 & 28.5 & 2.15 & 2.11  \\
    1e-06 & 60k  &  24.1 & 24.0 & 2.20 & 2.14  \\
    1e-06 & 82k  &  23.1 & 22.6 & 2.23 & 2.15   \\
    5e-06 & 15k  &  24.2 & 25.0 & 2.30 & 2.24   \\
    5e-06 & 60k  &  \textbf{21.9} & \textbf{22.0} & \textbf{2.34} & \textbf{2.25}   \\
    5e-06 & 73k  &  24.2 & 24.3 & 2.27 & 2.19   \\
    1e-05 & 15k  &  22.2 & 23.1 & 2.27 & 2.17   \\
    1e-05 & 60k  &  22.8 & \textbf{22.0} & 2.30 & 2.22   \\
    1e-05 & 40k  &  22.5 & 22.1 & 2.30 & 2.21   \\
    \bottomrule
    \end{tabular}
    \caption{Hyperparameter tuning of the TTS model: \WERR\ of synthetic male and female development and test sets, for various learning rate and training step combinations. \textit{None} indicates no fine-tuning was applied.}       \label{tab:hyperparameters_TTS}
\end{table}

\begin{table}[htb]
    \centering
    \begin{tabular}{llrrrr}
    \toprule
    Lower & Upper & \multicolumn{2}{c}{\WERR} & \multicolumn{2}{c}{\MOS}\\
    bound & bound & $\text{dev}$ & $\text{test}$ & $\text{dev}$ & $\text{test}$  \\
    \midrule
    None & None  &  21.9 & 22.0 & 2.34 & 2.24  \\ 
    0.5 & 1.5  &  \textbf{21.7} & 21.8 & 2.34 & \textbf{2.26}  \\
    0.7 & 1.5  &  21.8 & \textbf{21.3} & \textbf{2.35} & \textbf{2.26}    \\ 
    0.8 & 1.2  &  21.9 & 21.4 & \textbf{2.35} & \textbf{2.26}  \\
    
    \bottomrule
    \end{tabular}
    \caption{Filtering of generated data based on duration: {\%WER$_\text{R}$} and utMOS evaluation for various lower and upper bounds of the acceptable ratio range. \textit{None} indicates that no filtering was applied.}
    \label{tab:duration_TTS}
\end{table}

\begin{table*}[htb]
     \centering
     \begin{tabular}{llrrrrrr}
        \toprule
         &  & Speech dur. & N. utt. & N. words &  N. speakers & N. female & Dur. female \\
         Dataset &  Split & (h)         &         &          &              & (\%) & (\%) \\
        \midrule
        \multirow[t]{3}{*}{Bast} & Dev & 4.0 & 999 & 36.2K & 15 & 40.0 & 50.6 \\
         & Test & 8.0 & 2.24K & 74.1K & 19 & 42.1 & 49.0 \\
         & Train & 72.7 & 29.7K & 714K & 46 & 23.9 & 12.6 \\
        \cmidrule{1-8}
        \multirow[t]{3}{*}{Charb} & Dev & 5.0 & 1.53K & 49.5K & 31 & 48.4 & 49.0 \\
         & Test & 10.0 & 3.6K & 102K & 53 & 49.1 & 50.2 \\
         & Train & 709 & 301K & 7.28M & 339 & 19.8 & 21.9 \\
        \bottomrule
        \end{tabular}
     \caption{Commissions QC statistics.}         
     \label{tab:corpus_stat}
\end{table*}

\begin{table*}[htb]
    \centering
    \begin{tabular}{llrrrr}
    \toprule
     &  & \multicolumn{4}{c}{\WER} \\ \cmidrule{3-6}
    Train & Test & $\text{dev}_\text{F}$ & $\text{dev}_\text{M}$ & $\text{test}_\text{F}$ & $\text{test}_\text{M}$ \\
    \midrule
    \bastr & \bastr & 13.8 & 13.5 & 12.0 & 16.3  \\ 
    \bastr & {$\text{Bast}_\text{S-90,T-0.1}$} & 18.0 & 17.5 & 16.3 & 18.8  \\
    \bastr & {$\text{Bast}_\text{S-90,T-0.65}$} & 19.6 & 19.5 & 17.6 & 20.5  \\
    {$\text{Bast}_\text{S-90,T-0.1}$} & {$\text{Bast}_\text{S-90,T-0.1}$} & 11.1 & 11.0 & 10.4 & 10.9  \\
    {$\text{Bast}_\text{S-90,T-0.65}$} & {$\text{Bast}_\text{S-90,T-0.65}$} & 12.2 & 11.6 & 11.1 & 12.3 \\
    {$\text{Bast}_\text{S-90,T-0.1}$} & \bastr & 38.4 & 29.3 & 28.7 & 37.2  \\
    {$\text{Bast}_\text{S-90,T-0.65}$} & \bastr & 26.0 & 27.4 & 23.0 & 34.4  \\
    \bottomrule
    \end{tabular}
    \caption{\WER on male and female development and test sets when training with real or synthetic data and testing on real or synthetic data, and the generation temperature of XTTS is $0.1$ or $0.65$.}
    \label{tab:temperature_01_asr}
\end{table*}

\begin{table*}[htb]
    \centering
    \begin{tabular}{lllrrrr}
    \toprule
     & & & \multicolumn{4}{c}{\%WER\dn} \\ \cmidrule{4-7}
    Model & Status & Train & \devf & \devm & \testf & \testm  \\
    \midrule
    espnet-transformer & scratch & $\text{Bast}_\text{S-90,T-0.1}$ & 38.4 & 29.3 & 29.3 & 33.0 \\
    espnet-transformer & scratch & $\text{Bast}_\text{S-90,T-0.65}$ & 26.0 & 27.4 & 23.0 & 34.4 \\
    espnet-transformer & scratch & $\text{Gen-text}_\text{Mix-350/10}$ & 25.6 & 26.9 & 24.7 & 30.7 \\
    espnet-transformer & scratch & $\text{Bast}_\text{S-360}$ & 28.2 & 23.9 & 21.8 & 32.3\\
    espnet-transformer & scratch & $\text{Bast}_\text{Mix-350/10}$ & 24.1 & 25.0 & 22.5 & 27.9\\
    espnet-transformer & scratch & $\text{Bast}_\text{Mix-710/10}$ & 21.8 & 23.2 & 19.7 & 25.9\\
    espnet-transformer & scratch & $\text{Bast}_\text{Mix-330/30}$ & 21.2 & 22.2 & 20.3 & 24.1\\
    espnet-transformer & scratch & $\text{Bast}_\text{Mix-300/60}$ & 20.2 & 21.5 & 19.4 & 23.1\\
    espnet-transformer & scratch & $\text{Bast}_\text{Mix-710/60}$ & 19.8 & 21.2 & 18.2 & 22.8\\
    espnet-transformer & scratch & \bastr & \textbf{13.8} & \textbf{13.5} & \textbf{12.0} & \textbf{16.3} \\
    whisper-medium & fine-tuned & $\text{Bast}_\text{S-360}$ & 9.5 & 11.8 & 11.2 & 18.3   \\
    whisper-medium & fine-tuned & $\text{Bast}_\text{Mix-350/10}$ & 6.6 & 7.9 & 6.9 & 12.1   \\
    whisper-medium & fine-tuned & $\text{Bast}_\text{R-10}$ & 5.6 & 7.7 & 5.7 & 10.2   \\
    whisper-medium & fine-tuned & \bastr & \textbf{5.0} & \textbf{6.6} & \textbf{4.9} & \textbf{9.2}  \\
    \bottomrule
    \end{tabular}
    \caption{Overview of \%WER across synthetic, hybrid, and real datasets, trained and evaluated with different models on the real Bastarache development and test sets, with results reported for each gender.}
    \label{tab:overview_asr}
\end{table*}

%% file: main.bbl
\begin{thebibliography}{33}
\expandafter\ifx\csname natexlab\endcsname\relax\def\natexlab#1{#1}\fi

\bibitem[{Aks{\"e}nova et~al.(2022)Aks{\"e}nova, Chen, Chiu, Esch, Golik
  et~al.}]{aksenova_accented_2022}
Al{\"e}na Aks{\"e}nova, Zhehuai Chen, Chung-Cheng Chiu, Daan~van Esch, Pavel
  Golik, et~al. 2022.
\newblock \href {http://arxiv.org/abs/2205.08014} {Accented {Speech}
  {Recognition}: {Benchmarking}, {Pre}-training, and {Diverse} {Data}}.
\newblock ArXiv:2205.08014 [eess].

\bibitem[{Ardila et~al.(2020)Ardila, Branson, Davis, Henretty, Kohler
  et~al.}]{ardila_common_2020}
Rosana Ardila, Megan Branson, Kelly Davis, Michael Henretty, Michael Kohler,
  et~al. 2020.
\newblock Common voice: {A} massively-multilingual speech corpus.
\newblock In \emph{Proc. {LREC}}, pages 4218--4222.

\bibitem[{Banerjee et~al.(2024)Banerjee, Agarwal, and
  Ghosh}]{banerjee_high-precision_2024}
Sourav Banerjee, Ayushi Agarwal, and Promila Ghosh. 2024.
\newblock \href {http://arxiv.org/abs/2412.00055} {High-precision medical
  speech recognition through synthetic data and semantic correction:
  {UNITED}-{MEDASR}}.
\newblock ArXiv:2412.00055 [eess].

\bibitem[{Boyer et~al.(2021)Boyer, Shinohara, Ishii, Inaguma, and
  Watanabe}]{boyer2022studytransducerbasedendtoend}
Florian Boyer, Yusuke Shinohara, Takaaki Ishii, Hirofumi Inaguma, and Shinji
  Watanabe. 2021.
\newblock \href {https://ieeexplore.ieee.org/abstract/document/9688251} {A
  {{Study}} of {{Transducer Based End-to-End ASR}} with {{ESPnet}}:
  {{Architecture}}, {{Auxiliary Loss}} and {{Decoding Strategies}}}.
\newblock In \emph{Proc. {{ASRU}}}, pages 16--23.

\bibitem[{Bu et~al.(2017)Bu, Du, Na, Wu, and
  Zheng}]{bu2017aishell1opensourcemandarinspeech}
Hui Bu, Jiayu Du, Xingyu Na, Bengu Wu, and Hao Zheng. 2017.
\newblock \href {http://arxiv.org/abs/1709.05522} {{{AISHELL-1}}: {{An
  Open-Source Mandarin Speech Corpus}} and {{A Speech Recognition Baseline}}}.
\newblock In \emph{Proc. {{O-COCOSDA}}}.

\bibitem[{Canavan and Zipperlen(1996)}]{canavan_callfriend_1996}
Alexandra Canavan and George Zipperlen. 1996.
\newblock \href {https://catalog.ldc.upenn.edu/LDC96S48} {{CALLFRIEND}
  {Canadian} {French} {LDC96S48}}.
\newblock Linguistic Data Consortium.

\bibitem[{Casanova et~al.(2024)Casanova, Davis, Gölge, Göknar, Gulea, Hart,
  Aljafari, Meyer, Morais, Olayemi, and Weber}]{casanova_xtts_2024}
Edresson Casanova, Kelly Davis, Eren Gölge, Görkem Göknar, Iulian Gulea,
  Logan Hart, Aya Aljafari, Joshua Meyer, Reuben Morais, Samuel Olayemi, and
  Julian Weber. 2024.
\newblock \href {https://doi.org/10.48550/arXiv.2406.04904} {{XTTS}: a
  {Massively} {Multilingual} {Zero}-{Shot} {Text}-to-{Speech} {Model}}.
\newblock ArXiv:2406.04904 [eess].

\bibitem[{Fazel et~al.(2021)Fazel, Yang, Liu, Barra-Chicote, Meng, Maas, and
  Droppo}]{fazel_synthasr_2021}
Amin Fazel, Wei Yang, Yulan Liu, Roberto Barra-Chicote, Yixiong Meng, Roland
  Maas, and Jasha Droppo. 2021.
\newblock \href {https://doi.org/10.48550/arXiv.2106.07803} {{SynthASR}:
  {Unlocking} {Synthetic} {Data} for {Speech} {Recognition}}.
\newblock ArXiv:2106.07803 [cs].

\bibitem[{Hilmes et~al.(2024)Hilmes, Rossenbach, and
  Schlüter}]{hilmes_effect_2024}
Benedikt Hilmes, Nick Rossenbach, and {and}~Ralf Schlüter. 2024.
\newblock \href {https://doi.org/10.21437/SynData4GenAI.2024-10} {On the
  {Effect} of {Purely} {Synthetic} {Training} {Data} for {Different}
  {Automatic} {Speech} {Recognition} {Architectures}}.
\newblock In \emph{Synthetic {Data}’s {Transformative} {Role} in
  {Foundational} {Speech} {Models}}, pages 46--50.

\bibitem[{Hu et~al.(2022)Hu, Shen, Wallis, Allen-Zhu, Li, Wang, Wang, and
  Chen}]{hu2021loralowrankadaptationlarge}
Edward~J. Hu, Yelong Shen, Phillip Wallis, Zeyuan Allen-Zhu, Yuanzhi Li, Shean
  Wang, Lu~Wang, and Weizhu Chen. 2022.
\newblock \href {http://arxiv.org/abs/2106.09685} {Lora: Low-rank adaptation of
  large language models}.
\newblock In \emph{Proc. ICLR}.

\bibitem[{Kamble et~al.(2024)Kamble, Tathe, Kumbharkar, Bhandare, and
  Mitra}]{kamble_custom_2024}
Anand Kamble, Aniket Tathe, Suyash Kumbharkar, Atharva Bhandare, and Anirban~C.
  Mitra. 2024.
\newblock \href {http://arxiv.org/abs/2311.14836} {Custom {Data} {Augmentation}
  for low resource {ASR} using {Bark} and {Retrieval}-{Based} {Voice}
  {Conversion}}.
\newblock ArXiv:2311.14836 [cs, eess].

\bibitem[{Kim and Stern(2008)}]{kim_robust_2008}
Chanwoo Kim and Richard~M. Stern. 2008.
\newblock Robust signal-to-noise ratio estimation based on waveform amplitude
  distribution analysis.
\newblock In \emph{Proc. {Interspeech}}, pages 2598--2601.

\bibitem[{Kingma and Ba(2017)}]{kingma2017adammethodstochasticoptimization}
Diederik~P. Kingma and Jimmy Ba. 2017.
\newblock \href {http://arxiv.org/abs/1412.6980} {Adam: A method for stochastic
  optimization}.
\newblock In \emph{Proc. ICLR}.

\bibitem[{Koenecke et~al.(2020)Koenecke, Nam, Lake, Nudell, Quartey, Mengesha,
  Toups, Rickford, Jurafsky, and Goel}]{koenecke_racial_2020}
Allison Koenecke, Andrew Nam, Emily Lake, Joe Nudell, Minnie Quartey, Zion
  Mengesha, Connor Toups, John~R. Rickford, Dan Jurafsky, and Sharad Goel.
  2020.
\newblock Racial disparities in automated speech recognition.
\newblock \emph{Proceedings of the National Academy of Sciences},
  117(14):7684--7689.

\bibitem[{Laptev et~al.(2020)Laptev, Korostik, Svischev, Andrusenko,
  Medennikov, and Rybin}]{laptev_you_2020}
Aleksandr Laptev, Roman Korostik, Aleksey Svischev, Andrei Andrusenko, Ivan
  Medennikov, and Sergey Rybin. 2020.
\newblock \href {https://doi.org/10.1109/CISP-BMEI51763.2020.9263564} {You {Do}
  {Not} {Need} {More} {Data}: {Improving} {End}-{To}-{End} {Speech}
  {Recognition} by {Text}-{To}-{Speech} {Data} {Augmentation}}.
\newblock In \emph{Proc. {CISP}-{BMEI}}, pages 439--444.

\bibitem[{Le et~al.(2023)Le, Vyas, Shi, Karrer, Sari, Moritz, Williamson, Adi,
  Mahadeokar, and Hsu}]{le_voicebox_2023}
Matthew Le, Apoorv Vyas, Bowen Shi, Brian Karrer, Leda Sari, Rashel Moritz,
  Mary Williamson, Vimal Manohar~Yossi Adi, Jay Mahadeokar, and Wei-Ning Hsu.
  2023.
\newblock Voicebox: {{Text-Guided Multilingual Universal Speech Generation}} at
  {{Scale}}.
\newblock In \emph{Proc. {{NeurIPS}}}, volume~36, pages 14005--14034.

\bibitem[{Loshchilov and
  Hutter(2019)}]{loshchilov2019decoupledweightdecayregularization}
Ilya Loshchilov and Frank Hutter. 2019.
\newblock \href {http://arxiv.org/abs/1711.05101} {Decoupled {{Weight Decay
  Regularization}}}.
\newblock In \emph{Proc. {{ICLR}}}.

\bibitem[{Marcoux et~al.(2024)Marcoux, Richard, and
  Wolff}]{marcoux_estimation_2024}
Richard Marcoux, Laurent Richard, and Alexandre Wolff. 2024.
\newblock \href
  {https://www.odsef.fss.ulaval.ca/sites/odsef.fss.ulaval.ca/files/uploads/ODSEF_Estimation_Francophones_20241211.pdf}
  {Estimation des populations francophones dans le monde en 2024. {Sources} et
  démarches méthodologiques.}
\newblock Technical report, Université Laval, Observatoire démographique et
  statistique de l'espace francophone.

\bibitem[{Panayotov et~al.(2015)Panayotov, Chen, Povey, and
  Khudanpur}]{panayotov_librispeech_2015}
Vassil Panayotov, Guoguo Chen, Daniel Povey, and Sanjeev Khudanpur. 2015.
\newblock Librispeech: {An} {ASR} corpus based on public domain audio books.
\newblock In \emph{Proc. {ICASSP}}, pages 5206--5210.

\bibitem[{Park et~al.(2019)Park, Chan, Zhang, Chiu, Zoph, Cubuk, and
  Le}]{park_specaugment_2019}
Daniel~S. Park, William Chan, Yu~Zhang, Chung~Cheng Chiu, Barret Zoph, Ekin~D.
  Cubuk, and Quoc~V. Le. 2019.
\newblock {{SpecAugment}}: {{A Simple Data Augmentation Method}} for
  {{Automatic Speech Recognition}}.
\newblock In \emph{Proc. {{Interspeech}}}, pages 2613--2617.

\bibitem[{Pratap et~al.(2023)Pratap, Tjandra, Shi, Babu, Kundu
  et~al.}]{pratap2023a}
Vineel Pratap, Andros Tjandra, Bowen Shi, Paden Tomasello~Arun Babu, Sayani
  Kundu, et~al. 2023.
\newblock Scaling {Speech} {Technology} to 1,000+ {Languages}.
\newblock \emph{Journal of Machine Learning Research}, 25(97):1--52.

\bibitem[{Pratap et~al.(2020)Pratap, Xu, Sriram, Synnaeve, and
  Collobert}]{pratap_mls_2020}
Vineel Pratap, Qiantong Xu, Anuroop Sriram, Gabriel Synnaeve, and Ronan
  Collobert. 2020.
\newblock {MLS}: {A} large-scale multilingual dataset for speech research.
\newblock In \emph{Proc. {Interspeech}}, pages 2757--2761.

\bibitem[{Radford et~al.(2023{\natexlab{a}})Radford, Kim, Xu, Brockman,
  McLeavey, and Sutskever}]{radford_robust_2023}
Alec Radford, Jong~Wook Kim, Tao Xu, Greg Brockman, Christine McLeavey, and
  Ilya Sutskever. 2023{\natexlab{a}}.
\newblock \href {http://arxiv.org/abs/2212.04356} {Robust {Speech}
  {Recognition} via {Large}-{Scale} {Weak} {Supervision}}.
\newblock In \emph{Proc. {ICML}}, pages 28492--28518.

\bibitem[{Radford et~al.(2023{\natexlab{b}})Radford, Kim, Xu, Brockman,
  McLeavey, and Sutskever}]{radford2022robustspeechrecognitionlargescale}
Alec Radford, Jong~Wook Kim, Tao Xu, Greg Brockman, Christine McLeavey, and
  Ilya Sutskever. 2023{\natexlab{b}}.
\newblock Robust {{Speech Recognition}} via {{Large-Scale Weak Supervision}}.
\newblock In \emph{Proc. {{ICML}}}, pages 28492--28518.

\bibitem[{Rosenberg et~al.(2019)Rosenberg, Zhang, Ramabhadran, Jia, Moreno, Wu,
  and Wu}]{rosenberg_speech_2019}
Andrew Rosenberg, Yu~Zhang, Bhuvana Ramabhadran, Ye~Jia, Pedro Moreno, Yonghui
  Wu, and Zelin Wu. 2019.
\newblock \href {https://doi.org/10.1109/ASRU46091.2019.9003990} {Speech
  {Recognition} with {Augmented} {Synthesized} {Speech}}.
\newblock In \emph{Proc. {ASRU}}, pages 996--1002.

\bibitem[{Rousseau et~al.(2014)Rousseau, Boulianne, Deléglise, Estève, Gupta,
  and Meignier}]{rousseau_lium_2014}
Anthony Rousseau, Gilles Boulianne, Paul Deléglise, Yannick Estève, Vishwa
  Gupta, and Sylvain Meignier. 2014.
\newblock {LIUM} and {CRIM} {ASR} system combination for the {REPERE}
  evaluation campaign.
\newblock In \emph{Lecture {Notes} in {Computer} {Science}}, volume 8655 LNAI,
  pages 441--448.

\bibitem[{Saeki et~al.(2022)Saeki, Xin, Nakata, Koriyama, Takamichi, and
  Saruwatari}]{saeki_utmos_2022}
Takaaki Saeki, Detai Xin, Wataru Nakata, Tomoki Koriyama, Shinnosuke Takamichi,
  and Hiroshi Saruwatari. 2022.
\newblock \href {http://arxiv.org/abs/2204.02152} {{{UTMOS}}: {{UTokyo-SaruLab
  System}} for {{VoiceMOS Challenge}} 2022}.
\newblock In \emph{Proc. {{Interspeech}}}, pages 4521--4525.

\bibitem[{Sennrich et~al.(2016)Sennrich, Haddow, and
  Birch}]{sennrich_neural_2016}
Rico Sennrich, Barry Haddow, and Alexandra Birch. 2016.
\newblock \href {https://doi.org/10.48550/arXiv.1508.07909} {Neural {Machine}
  {Translation} of {Rare} {Words} with {Subword} {Units}}.
\newblock ArXiv:1508.07909.

\bibitem[{Serrand et~al.(2025)Serrand, Morsli, and Boulianne}]{serrand2025a}
Coralie Serrand, Amira Morsli, and Gilles Boulianne. 2025.
\newblock {CommissionsQC}: a {Qu{\'e}bec} {French} speech corpus for automatic
  speech recognition.
\newblock In \emph{Proc. {Interspeech}}, pages 3918--3922.

\bibitem[{Wang et~al.(2021)Wang, Riviere, Lee, Wu, Talnikar, Haziza,
  Williamson, Pino, and Dupoux}]{wang_voxpopuli_2021}
Changhan Wang, Morgane Riviere, Ann Lee, Anne Wu, Chaitanya Talnikar, Daniel
  Haziza, Mary Williamson, Juan Pino, and Emmanuel Dupoux. 2021.
\newblock {VoxPopuli}: {A} {Large}-{Scale} {Multilingual} {Speech} {Corpus} for
  {Representation} {Learning}, {Semi}-{Supervised} {Learning} and
  {Interpretation}.
\newblock In \emph{Proc. {ACL}}, pages 993--1003.

\bibitem[{Watanabe et~al.(2018)Watanabe, Hori, Karita, Hayashi, Nishitoba
  et~al.}]{watanabe_espnet_2018}
Shinji Watanabe, Takaaki Hori, Shigeki Karita, Tomoki Hayashi, Jiro Nishitoba,
  et~al. 2018.
\newblock {ESPnet}: {End}-to-{End} {Speech} {Processing} {Toolkit}.
\newblock In \emph{Proc. {Interspeech}}, pages 2207--2211.

\bibitem[{Zhang et~al.(2023)Zhang, Han, Qin, Wang, Bapna et~al.}]{zhang2023b}
Yu~Zhang, Wei Han, James Qin, Yongqiang Wang, Ankur Bapna, et~al. 2023.
\newblock \href {http://arxiv.org/abs/2303.01037} {Google {USM}: {Scaling}
  {Automatic} {Speech} {Recognition} {Beyond} 100 {Languages}}.
\newblock ArXiv:2303.01037 [cs, eess].

\bibitem[{Zheng et~al.(2021)Zheng, Liu, Gunceler, and
  Willett}]{zheng_using_2021}
Xianrui Zheng, Yulan Liu, Deniz Gunceler, and Daniel Willett. 2021.
\newblock \href {https://doi.org/10.48550/arXiv.2011.11564} {Using {Synthetic}
  {Audio} to {Improve} {The} {Recognition} of {Out}-{Of}-{Vocabulary} {Words}
  in {End}-{To}-{End} {ASR} {Systems}}.
\newblock In \emph{Proc. {ICASSP}}, pages 5674--5678.

\end{thebibliography}
